# Static Correlation and Dynamical Properties of Tb$^{3+}$-moments in Tb$_2$Ti$_2$O$_7$
# – Neutron Scattering Study –

Yukio Yasui[1,2], Masaki Kanada[1,2], Masafumi Ito[1,2], Hiroshi Harashina[1,2],
Masatoshi Sato[1,2], Hajime Okumura[3], Kazuhisa Kakurai[2,3*] and Hiroaki Kadowaki[4]

[1]*Department of Physics, Division of Material Science, Nagoya University,
Furo-cho, Chikusa-ku, Nagoya 464-8602, Japan*

[2]*CREST, Japan Science and Technology Corporation (JST)*

[3]*Neutron Scattering Laboratory, ISSP, The University of Tokyo,
106-1 Shirakata, Tokai, Ibaraki 319-1195, Japan*

[4]*Department of Physics, Tokyo Metropolitan University,
1-1 Minamiosawa, Hachioji, Tokyo 192-0397, Japan*

**Abstract**

Static and dynamical properties of the magnetic moment system of pyrochlore compound Tb$_2$Ti$_2$O$_7$ with strong magnetic frustration, have been investigated down to the temperature $T$=0.4 K by neutron scattering on a single crystal sample. The scattering vector (***Q***)-dependence of the magnetic scattering intensity becomes appreciable with decreasing $T$ at around 30 K, indicating the development of the magnetic correlation. From the observed energy profiles, the elastic, quasi elastic and inelastic components have been separately obtained. The quasi elastic component corresponds to the diffusive motion of the magnetic moments within the lowest states, which are formed of the lowest energy levels of Tb$^{3+}$ ions. Magnetic correlation pattern which can roughly reproduce the ***Q***-dependence of the scattering intensities of the elastic and quasi elastic component is discussed based on the trial calculations for clusters of 7 moments belonging to two corner-sharing tetrahedra. A possible origin of the glassy state, which develops at around 1.5 K with decreasing $T$ is discussed.

corresponding author : M. Sato (e-mail: e43247a@nucc.cc.nagoya-u.ac.jp)

*present address: Advance Science Research Center, JAERI, Tokai, Ibaraki 319-1195, Japan

§1. Introduction

R$_2$Ti$_2$O$_7$ (R=Y and various rare earth elements) belongs to the pyrochlore series of compounds with the general formula A$_2$B$_2$O$_7$, which has a cubic structure (space group F$d\bar{3}m$) and consists of two kinds of three-dimensional networks individually formed of corner-sharing tetrahedra of A$_4$ or B$_4$.[1] Due to this structural characteristics, magnetic moments at the A and/or B sites are expected to be frustrated, if their nearest neighbor interaction is antiferromagnetic. For classical Heisenberg spins with antiferromagnetic nearest neighbor interaction on the pyrochlore lattice, both mean-field theory[2] and Monte Carlo simulation[3] predict that there does not exist long-range magnetic order at finite temperature. The frustration is also expected even for the moment system of pyrochlore compounds with the ferromagnetic nearest neighbor interaction, if the moments have strong uniaxial anisotropy.[4-6]

For titanium pyrochlore compounds R$_2$Ti$_2$O$_7$, the existence of the magnetic order with the transition temperatures of 0.97 K, 1.25 K and 0.20 K has been reported for R=Gd, Er and Yb, respectively.[7,8] Tm$^{3+}$ ions in Tm$_2$Ti$_2$O$_7$ have the nonmagnetic ground state and the excitation energy to the triplet state is estimated to be 10.64 meV by magnetic susceptibility measurements and neutron inelastic scattering.[9] For R=Ho and Dy, the magnetic moments of R$^{3+}$ ions have the ferromagnetic nearest neighbor interaction and each magnetic moment at the corner of a tetrahedron lies along the principal axis parallel to the line which connects its site with the center of gravity of the tetrahedron, because the uniaxial anisotropy is very strong. In these compounds, the ground state of the moments within an isolated tetrahedron is given by the so-called "two-in and two-out" configuration, where two moments direct inwards and the other two direct outwards. By extending the "two-in and two-out" structure to the pyrochlore lattice, the ground state is expected to have macroscopic degeneracy and to be therefore highly frustrated. Because these moment configurations can be mapped to proton configurations of solid water, the system is called "spin ice."[4-6] It is noted, however, that the problem has been pointed out not to be strictly equivalent to that of solid water, because the dominant interaction among the moments has been found to be the long ranged dipole-dipole one.[10-12]

For Tb$_2$Ti$_2$O$_7$, the Weiss temperature estimated from the magnetic susceptibility data, is equal to be ~ 19 K,[13] indicating that nearest neighbor interaction between the Tb$^{3+}$ moments (~9.4 $\mu_B$) is antiferromagnetic. Figure 1 shows the magnetization (*M*) –magnetic field (*H*) curves of a single crystal of Tb$_2$Ti$_2$O$_7$ measured at 5 K in the applied field along three directions: [001], [110] and [111], where the anisotropy of the curves is found to be rather small as compared with that of spin ice system Ho$_2$Ti$_2$O$_7$, which exhibits significant anisotropic nature of the magnetization as shown in the inset of Fig. 1. Measurements of the muon spin-relaxation rate carried out for polycrystalline samples of Tb$_2$Ti$_2$O$_7$ have shown that magnetic ordering does not exist down to 70 mK,[13] which is much lower than the Weiss temperature (~19 K) stated above. Then, the effect of the geometrical frustration is significant in Tb$_2$Ti$_2$O$_7$. It is interesting to investigate in detail both static and dynamical magnetic properties of Tb$^{3+}$-moments of Tb$_2$Ti$_2$O$_7$.

In the present work, we have investigated magnetic properties of Tb$_2$Ti$_2$O$_7$ by means of neutron scattering on a single crystal specimen in the *T*-region between 0.4 K and 150 K. Taking the scattering data with much better energy resolution than that in the previous studies,[14-16)] we have found that there exist elastic, quasi elastic and inelastic components of the magnetic scattering and their intensities. Based on the data, the pattern of the possible short range correlation of the moments is discussed. We also discuss the detailed growth process of the magnetic correlation and the origin of freezing of the moments, which takes place at around 1.5 K with decreasing $T$.

§2. Experiments

A single crystal of $Tb_2Ti_2O_7$ with a volume of ~1 cm$^3$ was prepared by the floating zone (FZ) method as described in ref. 17. The magnetization was measured by using a SQUID magnetometer. Neutron measurements were carried out by using the triple axis spectrometer at C1-1 of the cold guide installed at JRR-3M of JAERI in Tokai. The 002 reflections of pyrolytic-graphite (PG) were used for both the monochromator and analyzer. The horizontal collimations in front of the sample were 30'-40' and a horizontally focusing analyzer was used. The final neutron energy $E_f$ was fixed at 2.424 meV or 3.098 meV, where the energy resolution $\Delta E$ (full width at half maximum) determined by the incoherent nuclear scattering from a vanadium rod was 46 $\mu$eV or 81 $\mu$eV, respectively. Higher-order neutrons were removed by a pyrolytic-graphite filter (and by a Be filter for the incident neutron energy $E_i \leq 5$ meV). The crystal was oriented with the [$hh$0] and [00$l$] axes in the scattering plane. The crystal was mounted in an Al can with He exchange gas which was attached to a liquid $^3$He or $^4$He cryostat.

§3. Experimental Results

The energy scan profiles of the magnetic scattering from the single crystal sample of $Tb_2Ti_2O_7$ were taken at various $Q$-points in the temperature region 0.4 K $\leq T \leq$ 50 K. Examples of the obtained profiles at 0.4 K are shown in Fig. 2. Figure 3 shows the energy profiles taken at $Q$=(0,0,2.1) at several temperatures. In both figures, we can see that the quasi elastic scattering component with asymmetric shape with respect to the $E$=0 point, is superposed onto the elastic component which has the Gaussian profile. (The asymmetry is just due to the Bose factor.) The intensity of the elastic scattering depends on the temperature and the scattering vector $Q$. It decreases with increasing magnetic field $H$,[15)] indicating that the magnetic scattering contribution exists in the elastic scattering.

Figures 4(a) and 4(b) show the energy scan profiles obtained at several temperatures with $\Delta E$ =81 $\mu$eV at $Q$=(1.25,1.25,0) and (1.9,1.9,1.9), respectively. Well-defined magnetic excitations are observed at the energy region of 1 meV ~ 2 meV. With decreasing $T$, the softening and the hardening of the excitation take place at around $Q$=(1.25,1.25,0) and (1.9,1.9,1.9), respectively, and the energy dispersion becomes significant. We have now found that the magnetic scattering has all the elastic, quasi elastic and inelastic contributions.

Figure 5 shows the scattering intensity map taken in the ($h,h,l$) plane in the reciprocal space at 0.4 K and at $E$=0 meV with $\Delta E$=81 $\mu$eV. The nuclear Bragg peaks observed at $Q$-points with even $h$ and $l$ or odd $h$ and $l$ are removed. It should be noted that the intensity is the sum of the magnetic elastic, a part of quasi elastic and the nuclear incoherent scattering. A strong intensity peak is observed in the region around $Q$= (0,0,2) and the region extends towards (1,1,1) and (1,1,3). Intensity peaks are also observed at $Q$=(0.5,0.5,0.5), (1.5,1.5,1.5), (1,1,0) and (1.7,1.7,0). Figures 6(a)-6(c) show the $Q$-dependence of the scattering intensity observed at several temperatures along (0,0,$l$), ($h$,$h$,2) and ($h$,$h$,0), respectively, at $E$=0 meV with $\Delta E$ =81 $\mu$eV. The $Q$-dependence becomes weak with increasing $T$ and almost disappears at $T$ ~ 50 K.

Figure 7 shows the $T$-dependence of the scattering intensities at $Q$=(0,0,2.1) and (2,2,1.4) measured at $E$=0 meV with $\Delta E$ =81 $\mu$eV. The scattering intensities increase with decreasing $T$ down to ~ 30 K in an almost $Q$-independent manner (see Figs. 6(a)-6(c)). Below ~ 30 K, the difference between the observed $T$-dependences at different $Q$-points becomes significant. It suggests that the short-range correlation begins to grow at around this temperature with decreasing $T$. With further decreasing $T$, a sharp upturn of the intensity is observed at $Q$=(0,0,2.1) at ~1.5 K. Hysteresis of the intensity-temperature curve at $Q$=(0,0,2.1) has been observed below ~ 1.5 K, when the temperature is changed up and down, as shown in the inset of Fig. 7. The curve depends on the cooling rate, too. Gardner *et al.*[13] reported, based on results of the magnetic susceptibility and μSR measurements, that the moment system remains paramagnetic down to 70 mK. However, the present results indicate that the moment system begins to freeze at around 1.5 K with decreasing $T$.

§4. Discussion

In the previous chapter, we have shown the existence of the elastic, quasi elastic and inelastic components of the magnetic scattering. Here, the intensity profiles of the elastic and quasi elastic components are first analyzed, where the dynamical structure factor $S(Q,\omega)$ is described as,

$$S(Q,\omega) = A\delta(\omega) + B\frac{1}{1-e^{-\beta\omega}} \cdot \chi_1(Q) \cdot \frac{\Gamma_1(Q)\cdot\omega}{\omega^2 + \Gamma_1(Q)^2}, \qquad (1)$$

where $\Gamma_1(Q)$ and $\chi_1(Q)$ are the broadening of the quasi elastic component and the static susceptibility, respectively. A and B are the scale factors and $\beta=(k_BT)^{-1}$. The first term corresponds to the elastic scattering and give the Gaussian profile of the observed intensity with the full width at half maximum of $\Delta E$ after the resolution convolution, and the second term represents the quasi elastic scattering, for which we have not made the resolution correction to obtain the intensity profile, because the width $\Gamma_1(Q)$ is much larger than the resolution width. In Figs. 2 and 3, the solid lines are the fitted curves, and the broken lines show the contribution of the quasi elastic component. The quality of the fitting is rather good, which indicates that the quasi elastic scattering exists in the profiles. The quasi elastic component could not be separated from the elastic one in the previous studies carried out by thermal neutrons with the energy resolution $\Delta E$=0.7 meV.[14-16]

Next, the intensity profiles of the inelastic scattering taken with $\Delta E$ =81 $\mu$eV in the energy region of 0.7 meV<$E$<3.0 meV are analyzed by using the following expression for the dynamical structure factor

$$S(\mathbf{Q},\omega) \propto \frac{1}{1-e^{-\beta\omega}} \cdot \chi_2(Q) \cdot \frac{\Gamma_2(Q) \cdot \omega}{(\omega-\omega_Q)^2 + \Gamma_2(Q)^2}, \qquad (2)$$

where $\omega_\mathbf{Q}$ and $\Gamma_2(\mathbf{Q})$ are the energy and the broadening of the inelastic components, respectively. The resolution correction was not made, because the resolution width ($\Delta E$=81 $\mu$eV) is much smaller than the value of $\Gamma_2(\mathbf{Q})$. The examples of the fits of eq. (2) to the experimental data are shown in Figs. 4(a) and 4(b) by solid curves. The fittings are satisfactory at least above 15 K. However, below 6 K, the quality of the fitting is not as good as that for $T$ higher than 15 K.

Figures 8(a) and 8(b) show the $\mathbf{Q}$-dependence of the integrated intensities of quasi elastic and elastic components at $T$=0.4 K along (0,0,$l$) and ($h,h,$2), respectively, where the integrated intensity of the quasi elastic component is approximately estimated by taking the summation of the intensity over the energy region of $-0.15$ meV $\leq E \leq 0.42$ meV. The scattering intensity taken at $E$=0 meV with $\Delta E$ =81 $\mu$eV is also shown. From these figures, the $\mathbf{Q}$-dependence of the integrated intensity of the quasi elastic component is found to be similar to that of the elastic one. Then, the quasi elastic component is considered to be the overdamped acoustic spin-wave-like excitation in the short-range ordered state.

Figures 9(a) and 9(b) show the examples of the dispersion relation along ($h,h,$0) and ($h,h,$2) obtained in the present analyses at 0.4 K, where the bottom of the dispersion curves are located at around the points $\mathbf{Q}$=(2$\pm\delta$,2$\pm\delta$,0) and ($\pm\delta,\pm\delta,$ 2) with $\delta\sim$0.75, and the top of the dispersion curves are located at around $\mathbf{Q}$=(2,2,2). These results are consistent with those obtained previously by thermal neutrons.[14,15] The $\mathbf{Q}$-dependence of $\omega_\mathbf{Q}$ does not have the clear correlation with the $\mathbf{Q}$-dependence of the intensities of the elastic and quasi elastic components.

Figure 10 shows the $T$-dependence of $\omega_\mathbf{Q}$ at several $\mathbf{Q}$-points. The energy of the excitation obtained at 50 K is almost $\mathbf{Q}$-independent, indicating that the excitation is regarded as a local crystal field excitation. However, because the energy softening or hardening becomes significant as the magnetic correlation develops, the inelastic component is considered to be an excitation, which has increasing collective nature with increasing magnetic correlation.[13]

Next, the short-range correlation pattern of the magnetic moments which can reproduce the $\mathbf{Q}$-dependence of the intensity of the elastic and quasi elastic components is discussed. We argued the correlation pattern in refs. 14-16, assuming that the inelastic component corresponds to the acoustic spin-wave-like excitation. However, we have clearly observed, with much better energy resolution than the previous one, the existence of the quasi elastic component, which can be considered to be the excitation within the states formed by the lowest levels of isolated $Tb^{3+}$ in the crystal field or the "overdamped" acoustic spin wave. The inelastic component therefore cannot be considered to correspond to the acoustic spin-wave-like excitation. Then, we have tried once more to search the correlation pattern which can reproduce the observed intensity map shown in Fig. 5 by calculating the square of the absolute static structure factor, $|F(\mathbf{Q})|^2$ ($\propto |\sum_j f(\mathbf{Q}) \mu_{j\perp} \times \exp(-i\mathbf{Q}\cdot\mathbf{r}_j)|^2$) for the cluster of 7 moments $\mu_j$ ($j = 1\sim 7$) at the corners of two tetrahedra in Fig. 11, where $f(\mathbf{Q})$ and $\mu_{j\perp}$ are the magnetic form factor and the perpendicular component of $\mu_j$ to the scattering vector $\mathbf{Q}$, respectively, and $\mathbf{r}_j$ indicates the position of $\mu_j$. The structure factor is averaged over the equivalent patterns, which can be obtained by exchanging the $x$, $y$ and $z$ directions.

We have, first, calculated $|F(\mathbf{Q})|^2$ for the two tetrahedra which satisfy the condition that the total magnetization of four moments within a tetrahedron is zero, and found that the local maxima of the intensity are located at $\mathbf{Q}=(1,1,1)$, $(1,1,3)$, $(2,2,0)$ and $(0,0,2)$ positions, which is quite different from the results in Fig. 5. Next, we try to consider the ground state configuration proposed in ref. 6 for antiferromagnetic pyrochlore systems with strong uniaxial anisotropy. The structure consists of alternation of layers of the tetrahedra, perpendicular to [100] direction with four moments direct inwards and four moments direct outwards as shown in Fig. 11(a), where each moment lies along the corresponding local principal axis. (The configuration is called "all-in or all-out" and observed in FeF$_3$.[18]) In this case, the degeneracy is broken, and there is a phase transition at finite temperature. The scattering intensity obtained for the structure has a maximum at $\mathbf{Q}=(2,2,0)$. It is equal to zero at $\mathbf{Q}=(0,0,2)$ and relatively large at $\mathbf{Q}=(1,1,3)$. These results indicate that the pattern shown in Fig. 11(a) does not explain the experimental results at all.

Palmer *et al.* proposed the ground state configuration of Heisenberg or *XY*-like moment system in antiferromagnetic pyrochlores with non-negligible dipole-dipole interaction. The structure is shown in Fig. 11(b), where the moments in each configuration are parallel to certain edges of the tetrahedron (The edges and the moments parallel to them are drawn with the same kinds of lines.). It is noted that all moments shown in Fig. 11(b) are perpendicular to the corresponding local principal axis. From the pattern, we have found that the maximum value of the intensity is located at $\mathbf{Q}=(0,0,2)$ which is similar to that of experimental results. However, the intensity map does not seem to satisfactorily reproduce the other characteristics.

Then, we have considered antiferromagnetic systems in which the uniaxial anisotropy energy and dipole-dipole interaction are comparable. After the calculation of the intensity for various possible patterns, we have found that the pattern derived by rotating the moments shown in Fig. 11(b) by 90 deg. or –90 deg. about the axes shown by the bold or broken lines gives a reasonable result. For the cluster of 7 moments, there are four distinct types of patterns which can be derived by the operations. Figures 11(c) and 11(c') show two of the correlation patterns, for examples. The intensity map obtained by averaging the intensities over these four types of patterns (In the actual calculation, the averaging is also carried out over the patterns derived from these four types of patterns by exchanging the *x*-, *y*- and *z*-directions). The result is shown in Fig. 11(d). (In the map, the lighter area indicates the higher intensity.) It can roughly reproduce the observed data. These patterns are almost degenerate within the rather small energy range, because only the dipole energies are different among the four types of patterns. Then, we may have to consider the state, which is described by the coexistence of these correlation patterns even at low temperatures.

The level scheme of the $4f$ electrons of $Tb^{3+}$ has been calculated here in the crystal field of the ions up to the fifth neighbor sites within the point charge approximation. The results indicate that both ground state and the first excited state are doublets approximately described by the eigen values of the $z$-component of the angular momentum, $J_z$ as $|J_z\rangle = |\pm 6\rangle$ and $|\pm 5\rangle$, respectively, while Gingras et al.[20] has reported based on their own model calculation that they are approximately described as $|\pm 4\rangle$ and $|\pm 5\rangle$, respectively. Although we do not have any experimental data to distinguish which of these results can describe the actual electronic structure, the difference of the $|J_z|$ values between the ground states and the excited states is unity in both cases. Because the energy dispersion of the inelastic scattering due to the excitation to the first excited state is, as observed in the present study, comparable to the energy difference (1 meV ~ 2 meV) between the ground states and the excited states, the magnetic interaction energy among the ions is strong enough to induce the significant mixing between the ground state and the first excited states. Then, the existence of the spin-wave-like excitation with the change of $|J_z|$ by 1 can be expected as the quasi elastic component within the states which are formed of the ground state levels of the $Tb^{3+}$ ions. The $M$-$H$ curves shown in Fig. 1 indicate the almost isotropic nature of the moments, which supports the possible existence of such the excitation above consideration. Gardner et al.[13,20] measured the crystal-field excitation spectra and pointed out that the second- and the third-excited state are located at 9.9 meV and 14.5 meV, respectively, which have the flat dispersion ($Q$-independent nature of the energies).

Figure 12(a) shows the $T$-dependence of the integrated intensities of the elastic component at $Q$=(0,0,2.1) and (1.5,1.5,1.9). Figures 12(b) and the inset show the $T$-dependence of the integrated intensities and the energy broadening $\Gamma_1(Q)$, respectively, of the quasi elastic component. The $Q$-dependence of the integrated intensity of the quasi elastic component becomes significant at ~ 30 K with decreasing $T$, indicating that the short-range correlation begins to grow at around this temperature, while the broadening $\Gamma_1(Q)$ does not exhibit significant $Q$-dependence from 50 K to 1.5 K. It does not exhibit significant $T$-dependence, either in the temperature region. At around 1.5 K, the integrated intensity of the elastic component exhibits a sharp increase with decreasing $T$, where the energy broadening of the quasi elastic component exhibits sharp decrease. The hysteretic and cooling-rate dependent behavior of the scattering intensity is also observed below this temperature (see Fig. 7). They are understood by the beginning of the moment-freezing.

Based on the results of the present studies, the behavior of the moments in $Tb_2Ti_2O_7$ can be understood as follows. At high temperatures, magnetic correlation does not exist and only the $Q$-independent magnetic scattering is observed and because the motion of the moments becomes slow gradually as $T$ decreases, the scattering intensity observed with the spectrometer setting with the zero transfer energy increases. The short-range correlation of the magnetic moments grows below ~ 30 K with decreasing $T$ and the intensity distribution shown in Figs. 5 and 6(a)-6(c) is realized at low temperatures.   However, the various magnetic correlation patterns are expected to coexist down to the temperature, at which these patterns become frozen and the glassy state is realized.

§5. Conclusion

The magnetic correlation and the dynamical properties of the frustrated pyrochlore antiferromagnet $Tb_2Ti_2O_7$ have been investigated by means of neutron scattering using a single crystal specimen in the $T$-region between 0.4 K and 150 K, where the elastic, quasi elastic and inelastic components are separated.   Magnetic correlation pattern, which can roughly reproduce the $Q$-dependence of the scattering intensity of the elastic and quasi elastic components, is proposed by calculating the structure factor of the magnetic scattering for the clusters of 7 moments.   It seems that there exist many correlation patterns including the proposed pattern with almost degenerate energies.   The short-range correlation becomes appreciable with decreasing $T$ at around 30 K.   At the temperature, many correlation patterns (at least four distinct types of patterns and equivalent ones derived from these patterns by exchanging the $x$-, $y$- and $z$-directions) coexist.   With further decreasing $T$, these patterns become frozen and the glassy state develops at around 1.5 K before reaching the moment configuration of the lowest energy state.

Acknowledgements

Two of the authors (Y. Y and M. K.) were supported by a Research Fellowship of the Japan Society for the Promotion of Science for Yong Scientists.

References

1)  for example, M. A. Subramanian, G. Aravamudan and G. V. Subba Rao: Prog. Solid State Chem. **15** (1983) 55.
2)  J. N. Reimers, A. J. Berlinsky and A. C. Shi: Phys. Rev. B **43** (1991) 865.
3)  J. N. Reimers: Phys. Rev. B **45** (1992) 7287.
4)  A. P. Ramirez, A. Hayashi, R. J. Cava, R. Siddharthan and B. S. Shastry: Nature **399** (1999) 333.
5)  M. J. Harris, S. T. Bramwell, D. F. McMorrow, T. Zeiske and K. W. Godfley: Phys. Rev. Lett. **79** (1997) 2554.
6)  S. T. Bramwell and M. J. Harris: J. Phys.: Condens. Matter **10** (1998) L215.
7)  N. P. Raju, M. Dion, M. J. P. Gingras, T. E. Mason and J. E. Greedan: Phys. Rev. B **59** (1999) 14489.
8)  H. W. J. Blote, R. F. Wielinga and W. J. Huiskamp: Physica **43** (1969) 549.
9)  M. P. Zinkin, M. J. Harris, Z. Tun, R. A. Cowley and B. M. Wanklyn: J. Phys.: Condens. Matter **8** (1996) 193.
10) B. C. den Hertog and M. J. P. Gingras: Phys. Rev. Lett. **84** (2000) 3430.
11) S. T. Bramwell, M. J. Harris, B. C. den Hertog. M. J. P. Gingras, D. F. McMorrow, A. R. Wildes, A. Cornelius, J. D. M. Champion, R. G. Melko and T. Fennell: Phys. Rev. Lett. **87** (2001) 047205.
12) M. Kanada, Y. Yasui, Y. Kondo, S. Iikubo, M. Ito, H. Harashina, M. Sato, H. Okumura, K. Kakurai and H. Kadowaki: submitted to J. Phys. Soc. Jpn.
13) J. S. Gardner, S. R. Dunsiger, B.D. Gaulin, M. J. P Gingras, J. E. Greedan, R. F. Kiefl, M. D. Lumsden, W. A. MacFarlane, N. P. Raju, J. E. Sonier, I. Swainson and Z. Tun: Phys. Rev. Lett. **82** (1999) 1012.
14) M. Kanada, Y. Yasui, M. Ito, H. Harashina, M. Sato, H. Oakumura and K. Kakurai: J. Phys. Soc. Jpn. **68** (1999) 3802.
15) Y. Yasui, M. Kanada, M. Ito, H. Harashina, M. Sato, H. Okumura and K. Kakurai; J. Phys. Chem. Solids **62** (2001) 343.
16) M. Kanada, Y. Yasui, M. Ito, H. Harashina, M. Sato, H. Okumura, K. Kakurai and H. Kadowaki: Proc. 1st Int. Symp. Advanced Science Research, Advances in neutron Scattering Research, J. Phys. Soc. Jpn. **70** (2001) Suppl. A, p.200.
17) J. S. Gardner, B. D. Gaulin and D. McK. Paul: J. Crystal Growth **191** (1998) 740.
18) G. Ferey, R. De Pape, M. Leblanc and J. Pannetier: Rev. Chim. Miner. **23** (1986) 474.
19) S. E. Palmer and J. T. Chalker: Phys. Rev. B **62** (2000) 488.
20) M. J. P. Gingras, B. C. den Hertog, M. Faucher, J. S. Gardner, S. R. Dunsiger, L. J. Chang, B. D. Gaulin, N. P. Raju and J. E. Greedan: Phys. Rev. B **62** (2000) 6496.

Figure captions

Fig. 1    Magnetization $M$ of $Tb_2Ti_2O_7$ measured at 5 K along the three directions, [001], [110] and [111] are shown against the applied magnetic field $H$. Inset shows $M$-$H$ curves of spin ice system $Ho_2Ti_2O_7$ measured at 5 K in the applied field along [001], [110] and [111] directions.

Fig. 2    Energy profiles taken at 0.4 K at several $Q$-points with $\Delta E$=46 $\mu$eV, where solid lines show the results of the fitting to the total intensity. The broken lines show the intensities of the quasi elastic components. See text for details.

Fig. 3    Energy profiles taken at $Q$=(0,0,2.1) at several temperatures with $\Delta E$=46 $\mu$eV, where solid lines show the results of the fitting to the total intensity. The intensities of the quasi elastic components are shown by the broken lines.

Fig. 4    Energy profiles taken at several temperatures at (a)$Q$=(1.25,1.25,0) and (b)(1.9,1.9,1.9), respectively, where the solid lines show the results of the fitting.

Fig. 5    Map of the scattering intensity observed in the ($h,h,l$) plane in the reciprocal space at 0.4 K with the neutron transfer energy $E$=0 meV and $\Delta E$=81 $\mu$eV.

Fig. 6    $Q$-dependence of the scattering intensity taken at several temperatures with $E$=0 meV and $\Delta E$= 81 $\mu$eV is shown along (a)(0,0,$l$), (b)($h,h$,2) and (c)($h,h$,0), respectively.

Fig. 7    Temperature dependence of the scattering intensity measured at $Q$=(0,0,2.1) and (2,2,1.4) with $E$=0 meV and $\Delta E$=81 $\mu$eV. The inset shows the intensity-temperature curves on warming and cooling, where the curves on cooling are obtained at different cooling rate of 0.2 K/min. (closed circles) and 0.03 K/min. (open squares), respectively.

Fig. 8    $Q$-dependence of the integrated intensities of the quasi elastic and elastic components at $T$=0.4 K is shown along (a)(0,0,$l$) and (b)($h,h$,2), respectively, together with the scattering intensity taken with $E$=0 meV and $\Delta E$=81 $\mu$eV.

Fig. 9    The dispersion curves of the inelastic component are shown along (a)(0,0,$l$) and (b)($h,h$,2), respectively, at 04 K.

Fig. 10   Temperature dependence of the energy $\omega_Q$ of inelastic component at several $Q$-points in the reciprocal space. Broken lines are guides for the eye.

Fig. 11   (a)-(c') Schematic figures of the correlation patterns described in the text. (d)Map of the intensity averaged over the four types of magnetic patterns, two of which are shown in (c) and (c') (In the actual calculation, the averaging is also carried out over the patterns derived from these four types of patterns by exchanging the $x$-, $y$- and $z$-directions). Lighter area indicates the higher intensity in the map.

Fig. 12   Temperature dependence of the integrated intensities of (a)the elastic and (b)the quasi elastic scattering, at $Q$=(0,0,2.1) and (1.5,1.5,1.9). Broken lines are guides for the eye. In the inset of (b), $\Gamma_1(Q)$ is plotted against $T$ at $Q$=(0,0,2.1) and (1.5,1.5,1.9).

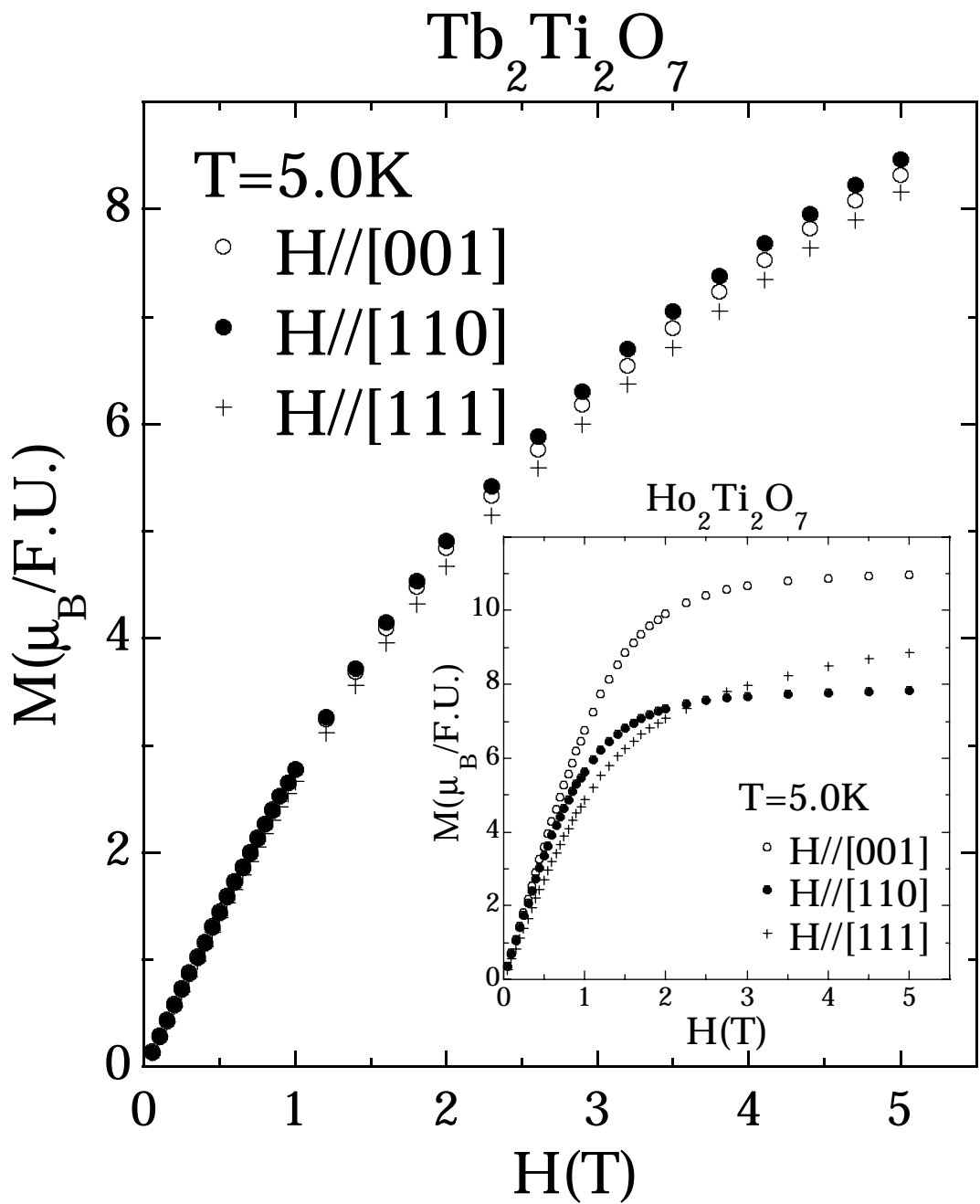

Fig. 1

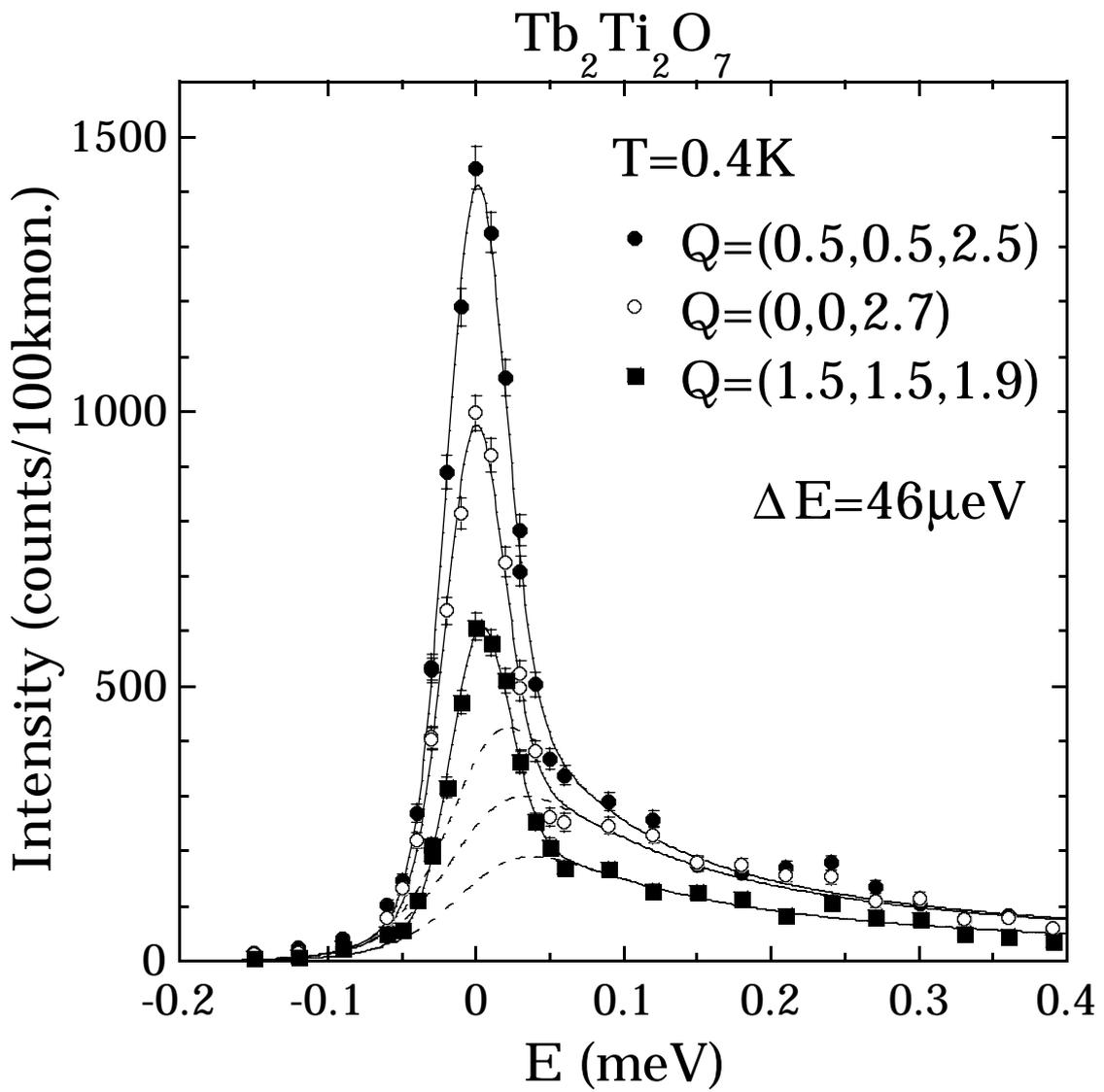

Fig.2

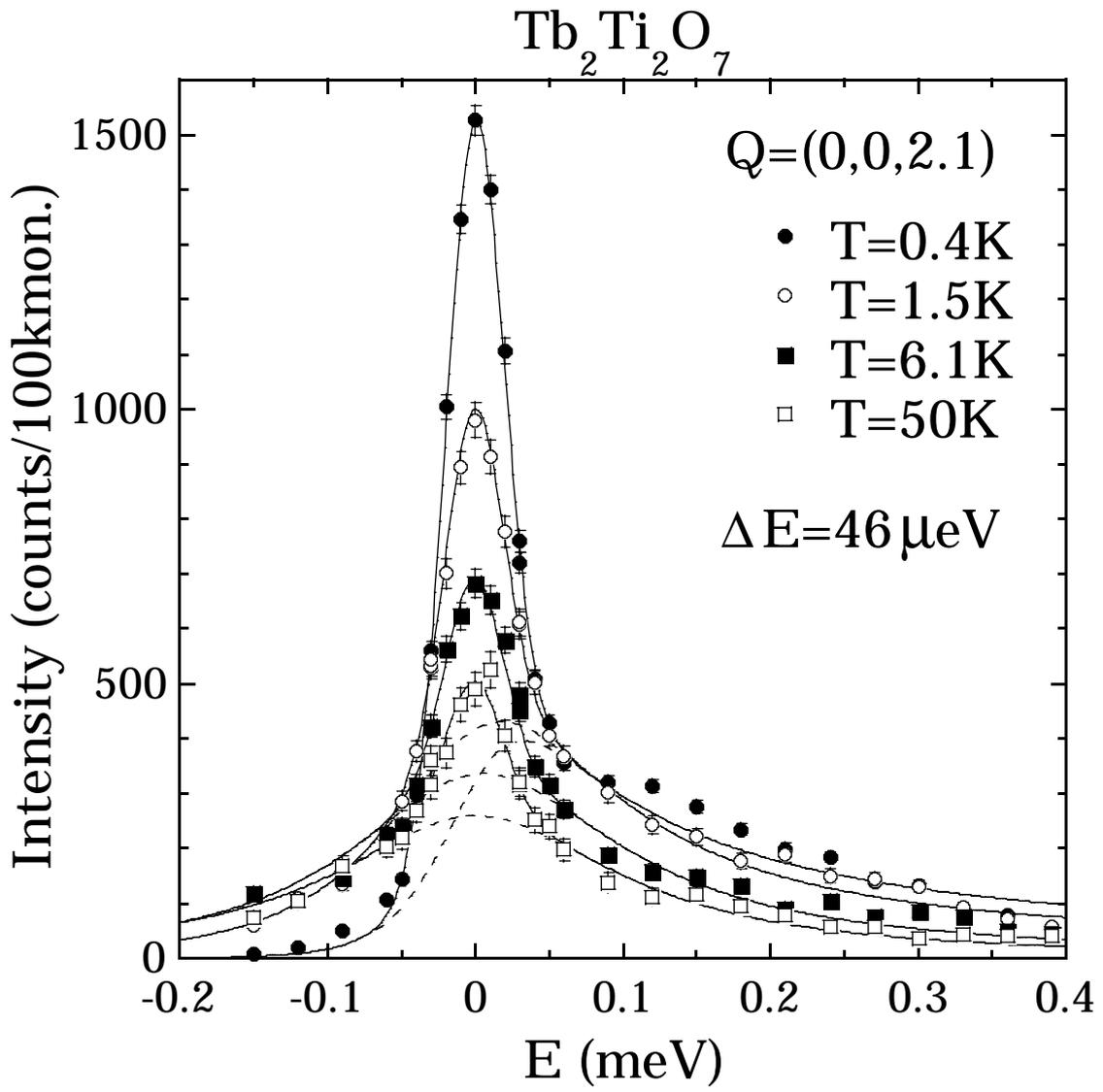

Fig. 3

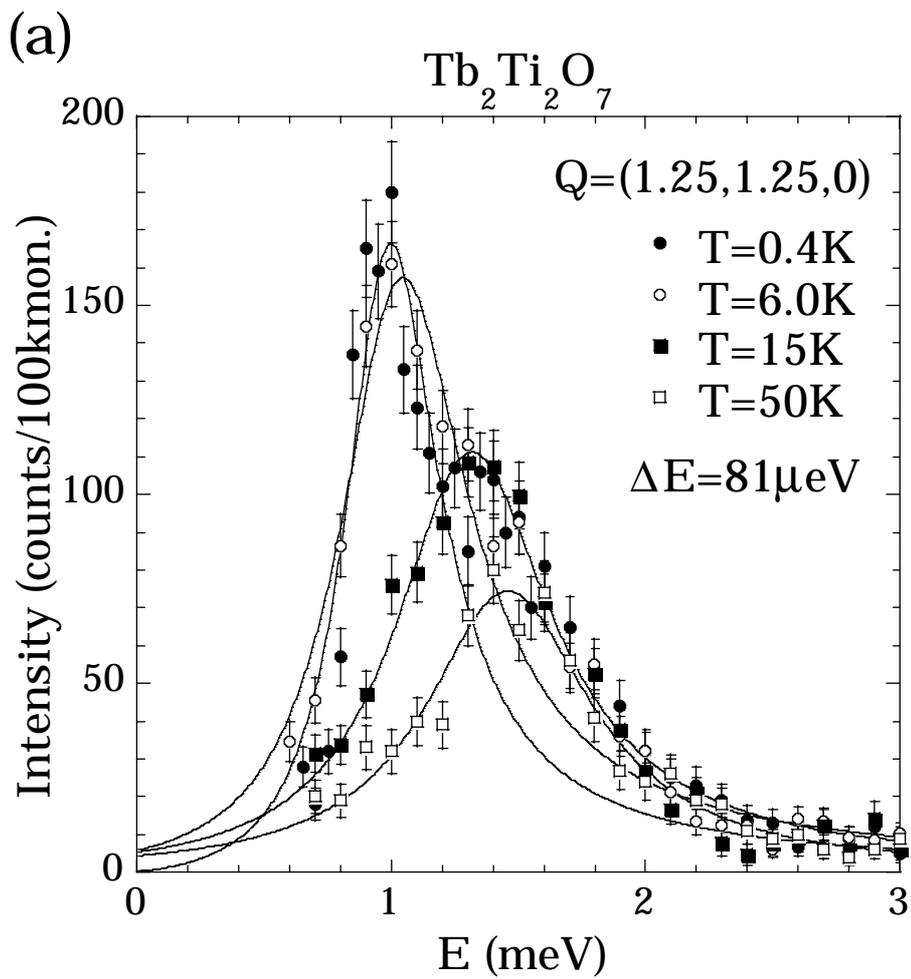

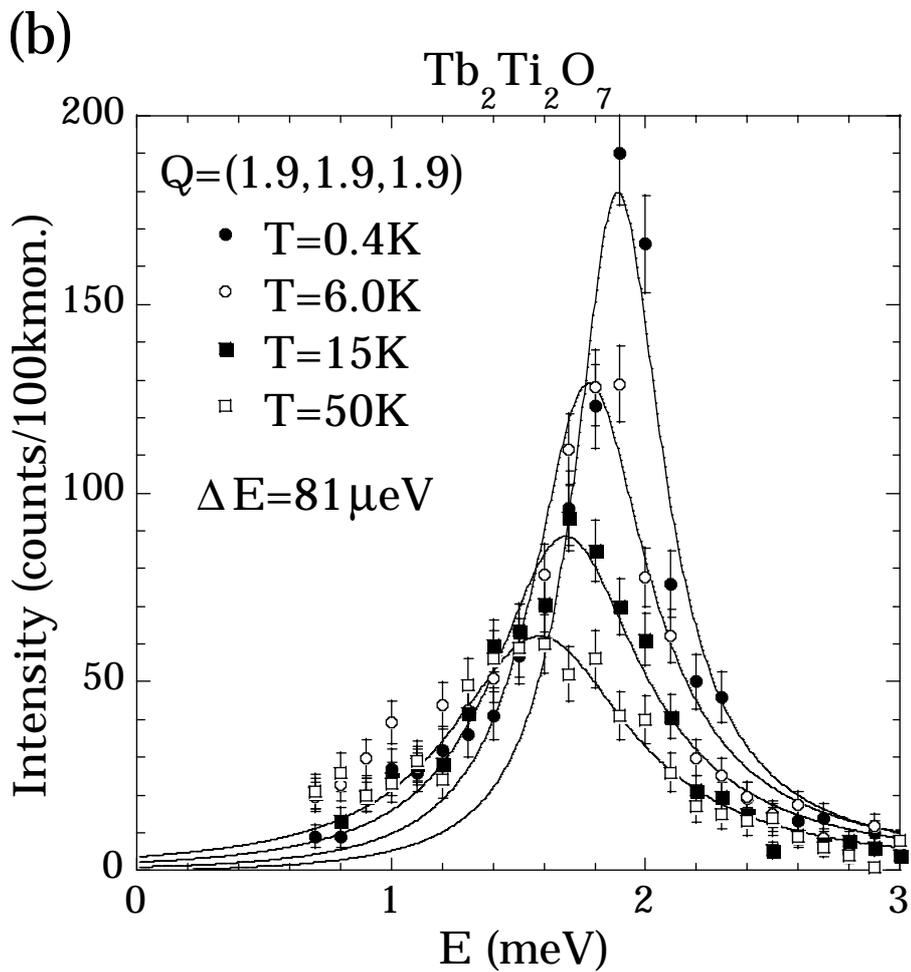

Fig. 4

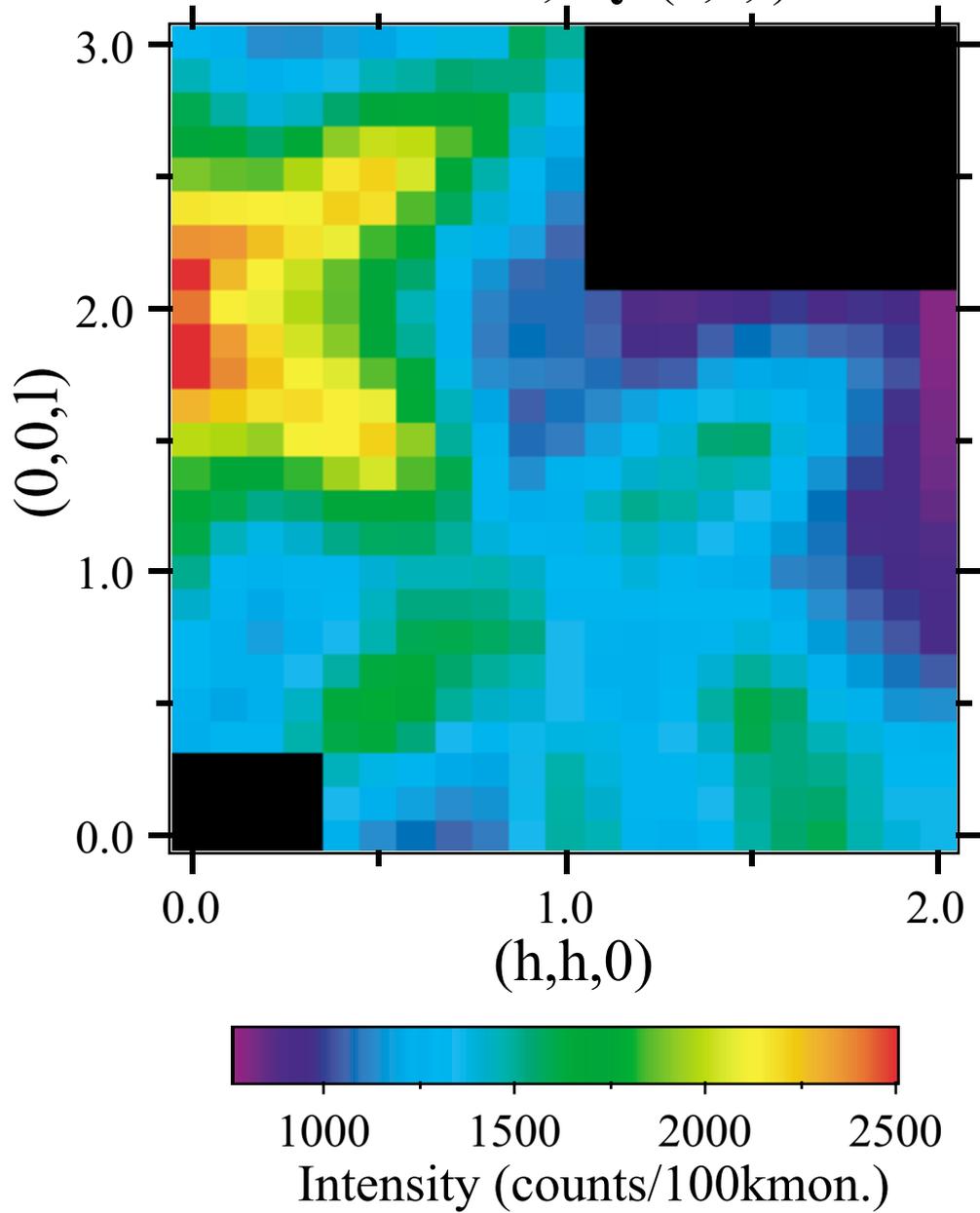

Fig. 5

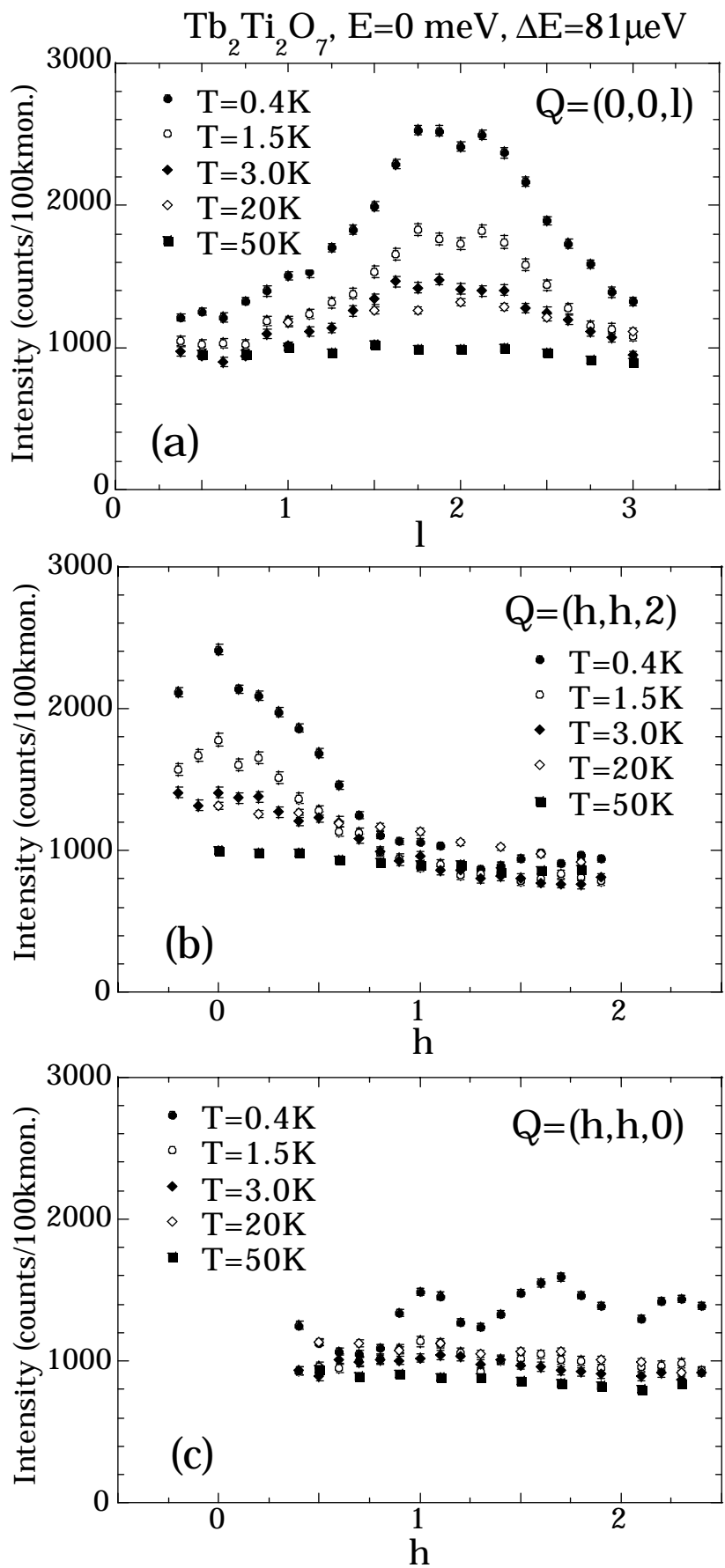

Fig. 6

Tb$_2$Ti$_2$O$_7$, E=0 meV, ΔE=81 μeV

Main plot labels: Q=(0,0,2.1); Q=(2,2,1.4)

Inset: Q=(0,0,2.1)
- on warming (0.03K/min.)
- on cooling (0.03K/min.)
- on cooling (0.2K/min.)

Fig. 7

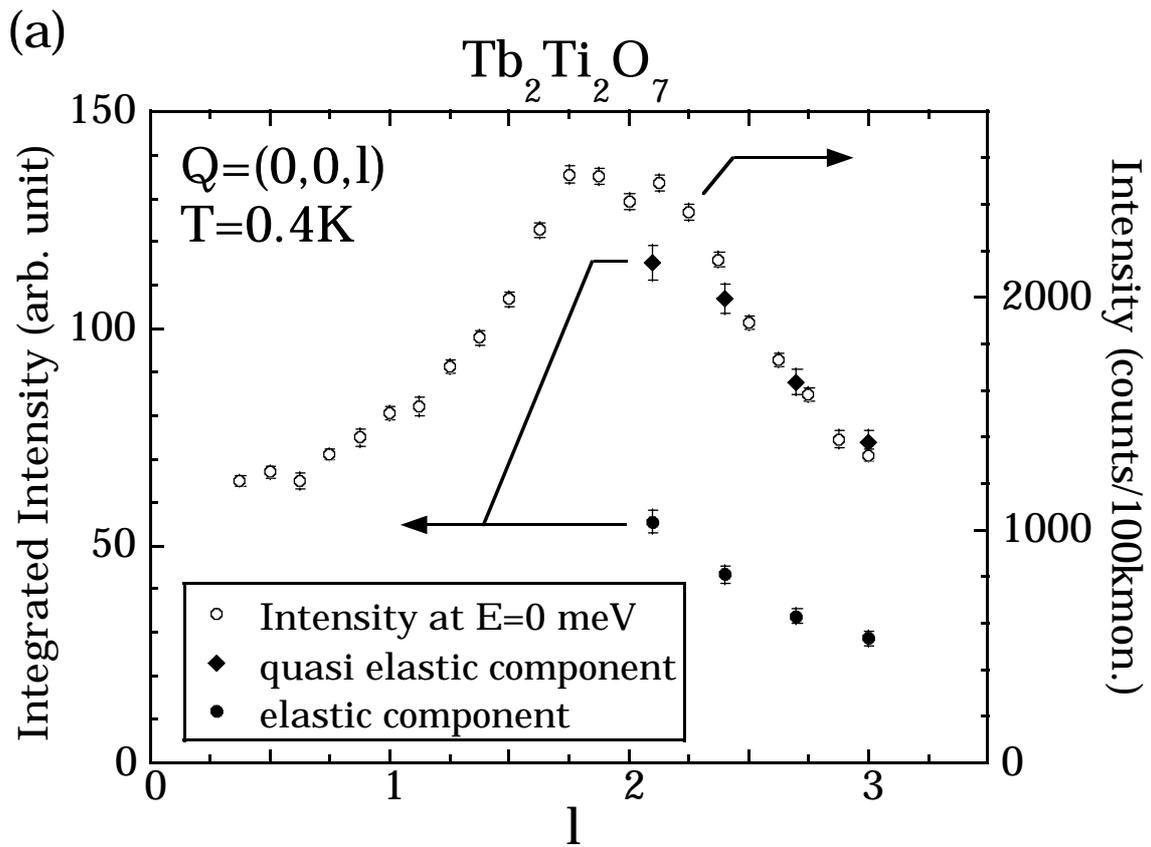

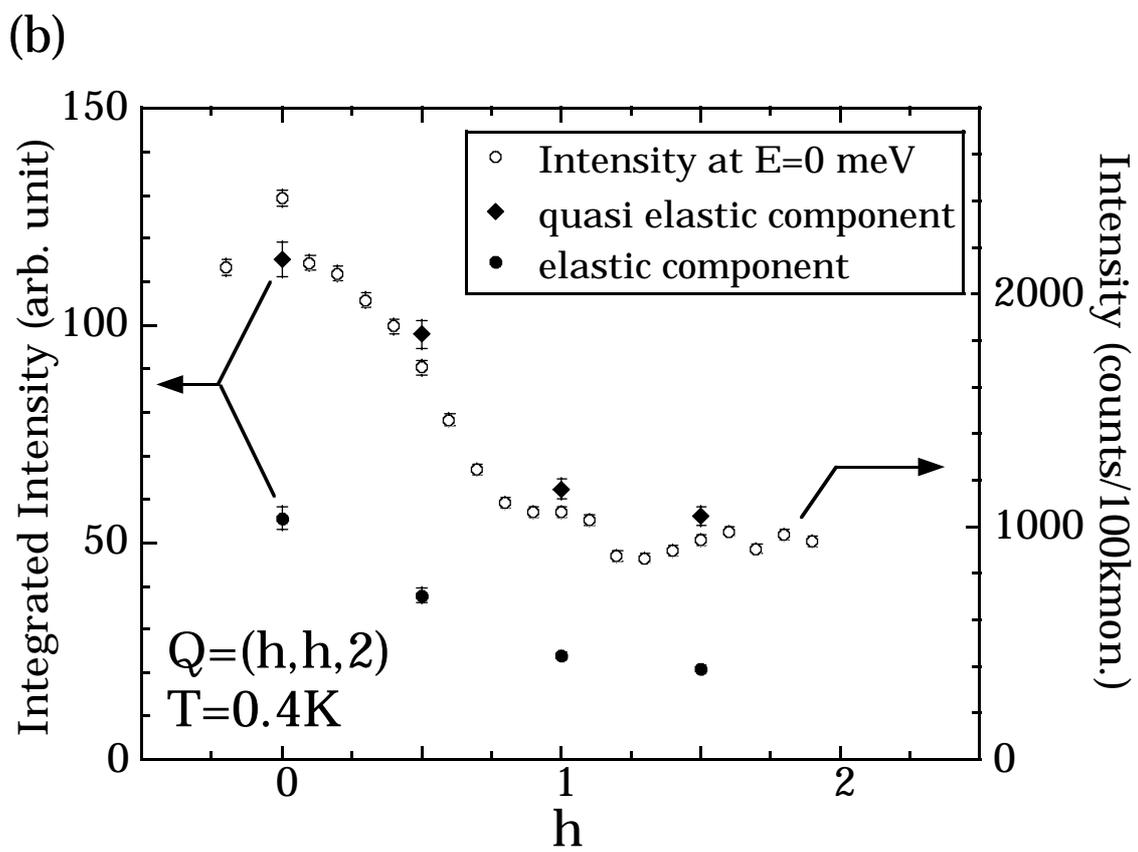

Fig. 8

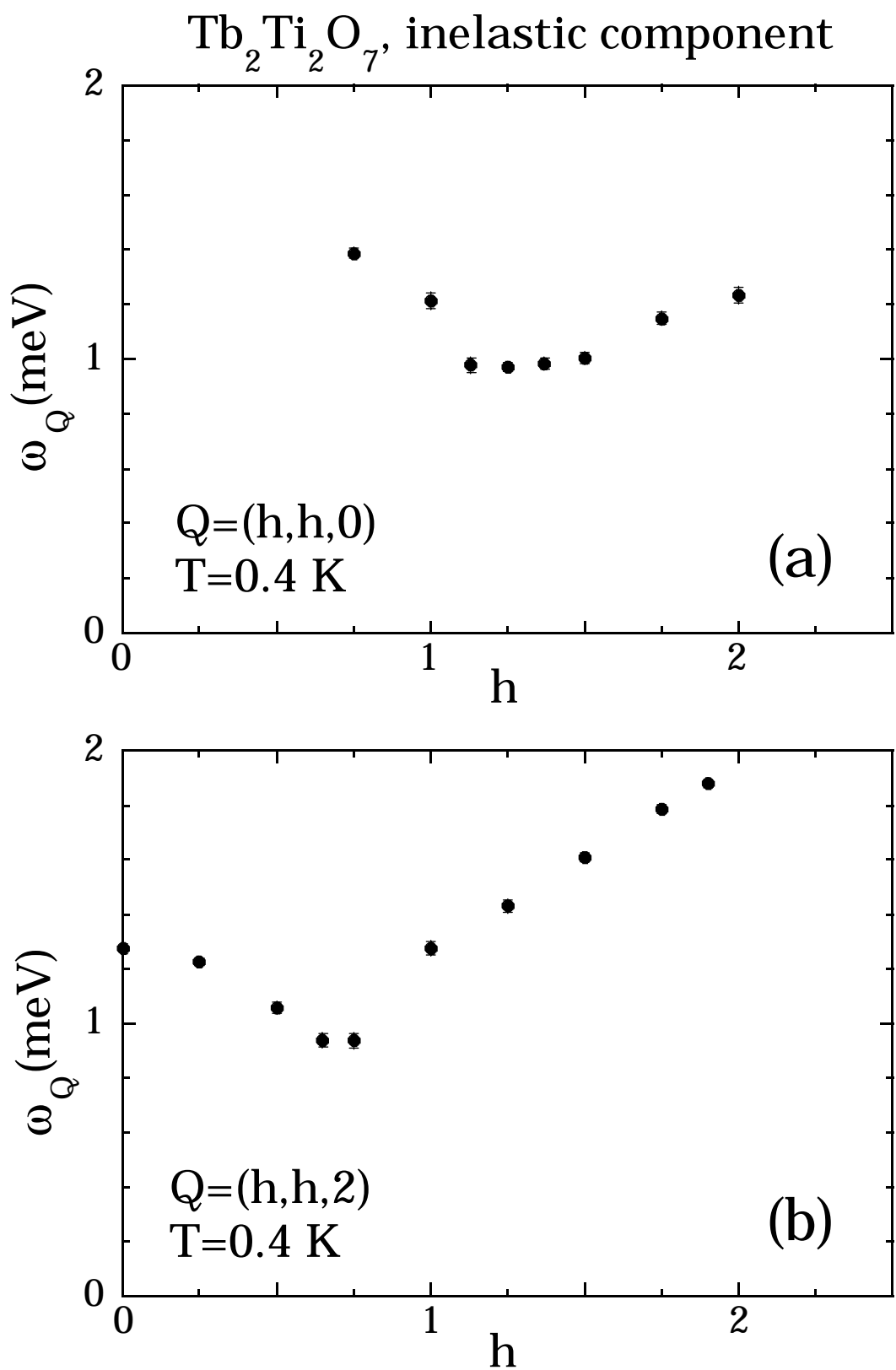

Fig. 9

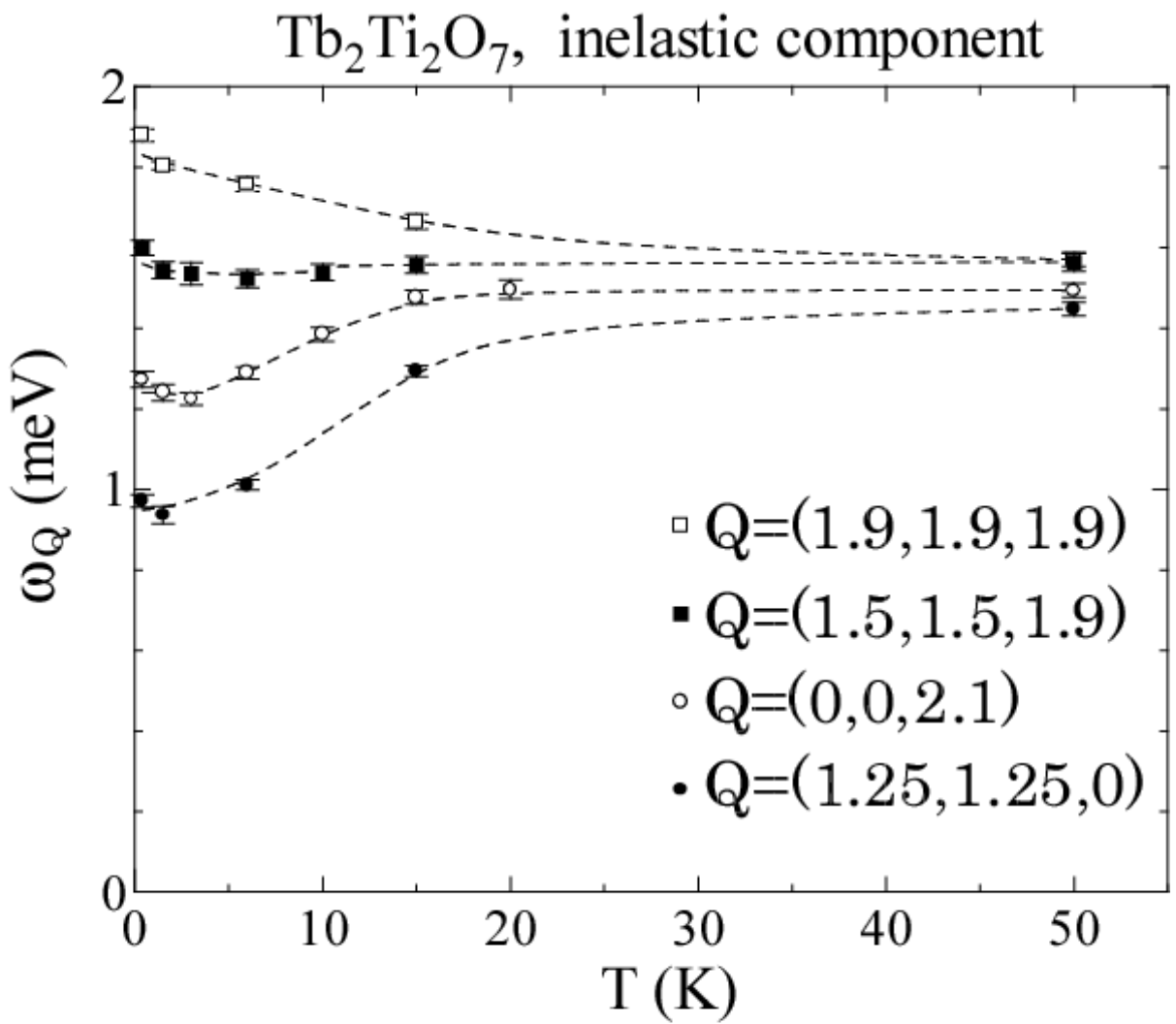

Fig. 10

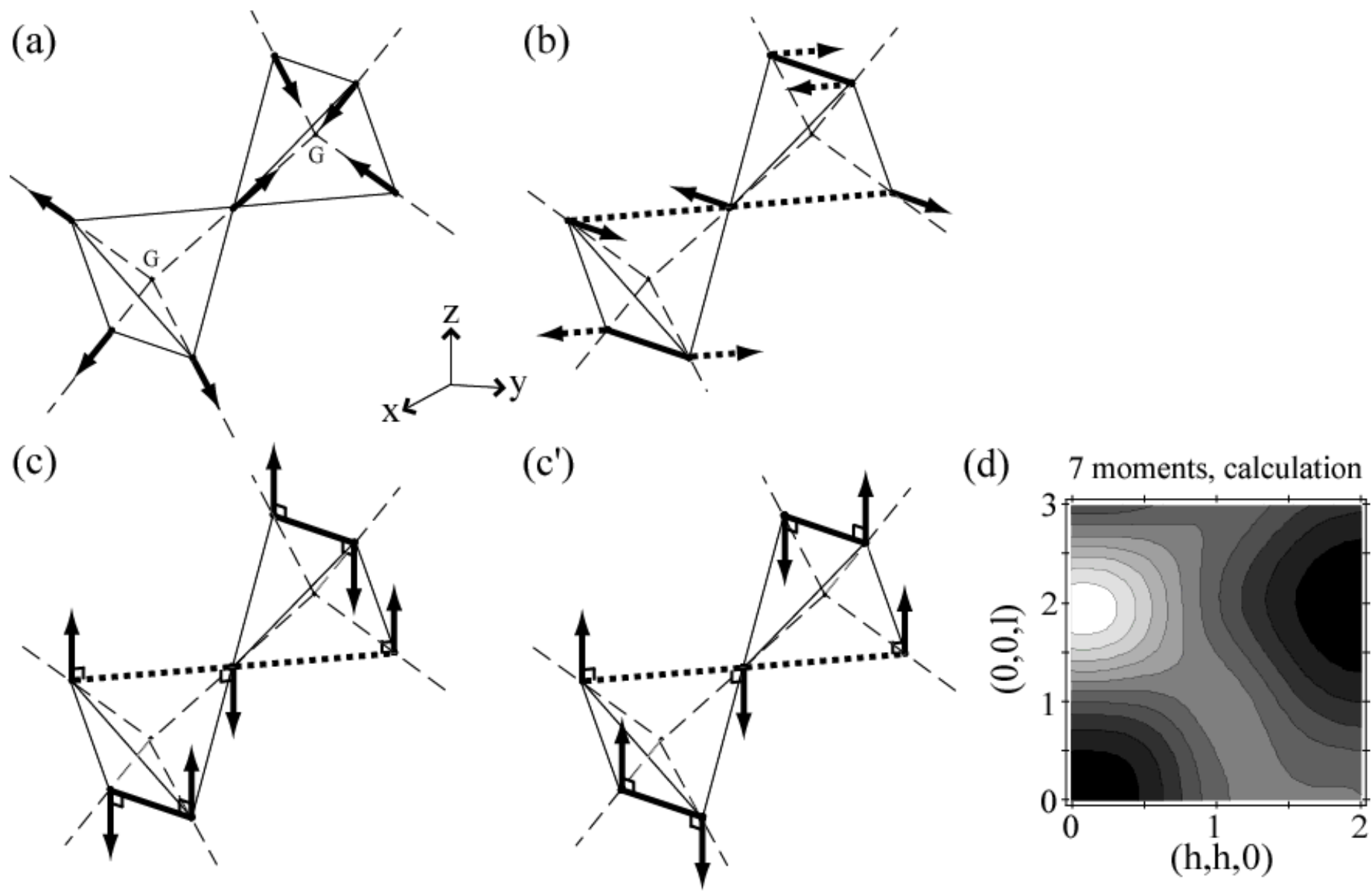

Fig. 11

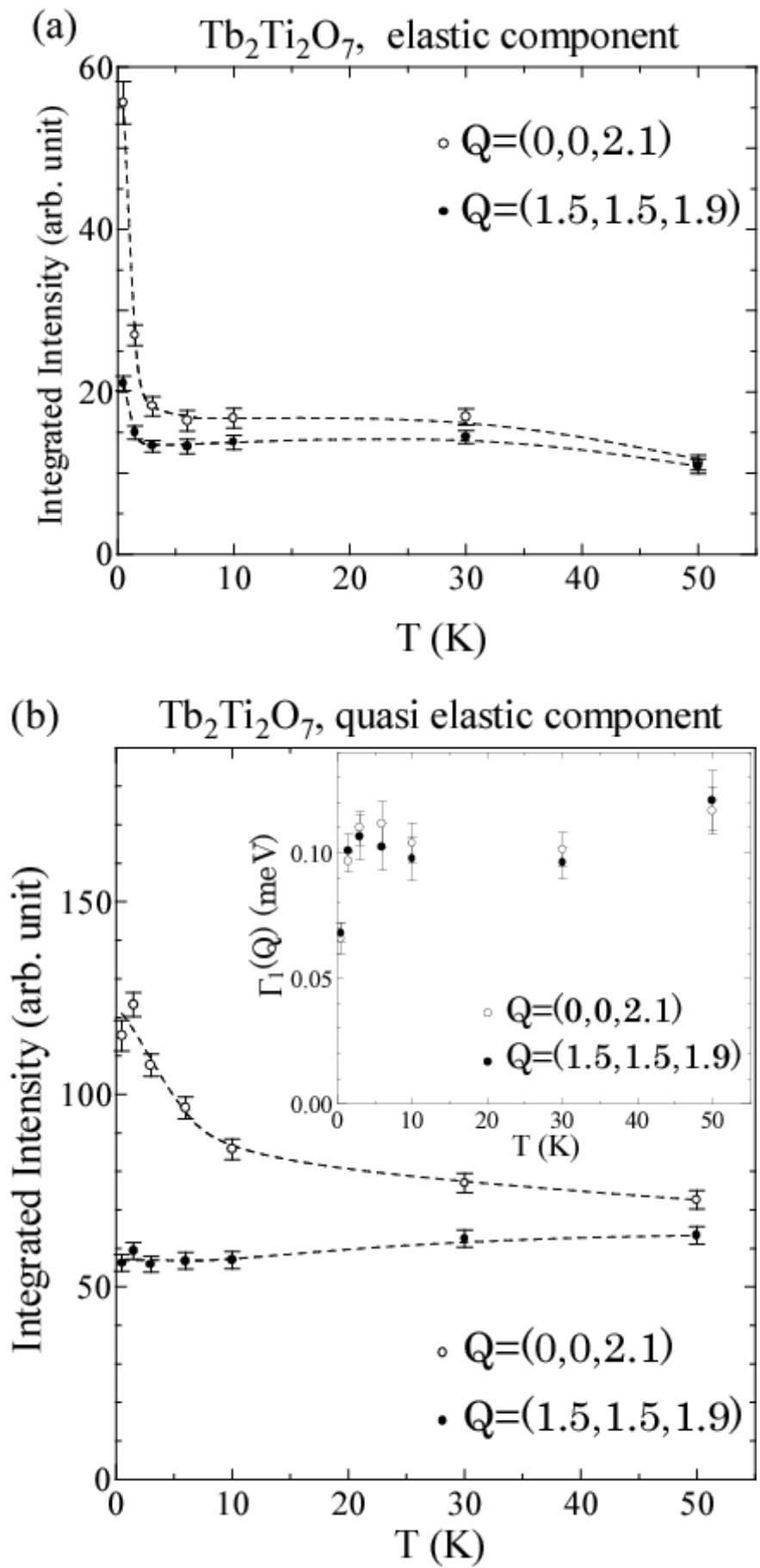

Fig. 12